\documentclass[twoside]{reporteq}
\usepackage{svcon2e}

\usepackage{latexsym}
\usepackage{amsmath}   %needed for \begin{align}... \end{align} environment
\usepackage{amsfonts}
\usepackage{amssymb}
\usepackage{graphicx}
\usepackage{amsthm}
\input{psfig}
\newcommand{\BC}{\mathbb{C}}
\newcommand{\BR}{\mathbb{R}}

\newcommand{\A}{{\cal A}}

\begin{document}

\chapter{Spinors in Quantum Geometrical Theory}

\chapterauthors{Daniel C.Galehouse}
%%%%if used by the author(s):
{ {\renewcommand{\thefootnote}{\fnsymbol{footnote}}
\footnotetext{\kern-15.3pt AMS Subject Classification: 
15A66, 81Q99, 81S99, 81V22, 83E15}
}}

\begin{abstract}

Spinors have played an essential but enigmatic role in modern physics
since their discovery.  Now that quantum-gravitational theories have
started to become available, the inclusion of a description of spin in
the development is natural and may bring about a profound
understanding of the mathematical structure of fundamental physics.  A
program to attempt this is laid out here.  Concepts from a known
quantum-geometrical~\index{quantum-geometrical theory} theory are
reviewed: (1) Classical physics is replaced by a suitable geometry as
a fundamental starting point for quantum mechanics.  (2) In this
context, a resolution is found for the enigma of wave-particle
duality.  (3) It is shown how to couple the quantum density to the
geometrical density.  (4) The mechanical gauge is introduced to allow
dimensional reduction.  (5) Absolute geometrical equivalence is
enforced. The concordant five-dimensional quantum-geometrical theory
is summarized to provide an orderly basis for the introduction of
spinors.  It is supposed that the Pauli--Dirac theory is adaptable. A
search is begun for a description that will generate spinors as a
natural tangent space.  Interactions other than gravity and
electrodynamics should then appear intrinsically.  These are
conjectured to be weak effects for electrons.

\noindent {\bf Keywords: }\par Dirac spinors, 
quantum-gravitational theory, geometry
\end{abstract}

\pagestyle{myheadings}
\markboth{Daniel Galehouse}{Spinors in Quantum Geometrical Theory}

\section{Introduction}

A study of Cartan's~\cite{ecartan} book suggests that spinors may
function at a fundamental level in the geometry of physics.  Interest
has continued unabated following discovery of the relativistic
equation of the electron by Dirac.  If applications are included, the
available literature is perhaps the largest of any single topic in
physics. Within the context of earlier studies by the
author,~\cite{bjga}, the contribution of fundamental spin to quantum
geometrical theory must be reevaluated

\begin{figure}[ht]
\centerline{\psfig{figure= 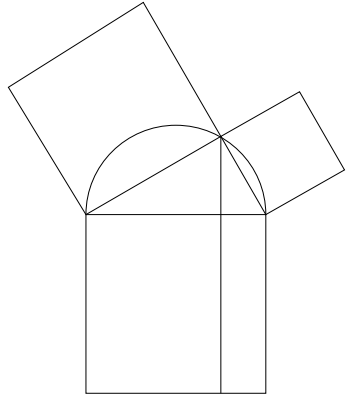,height=5.25cm,width=4.5cm}}
\caption{The fundamental concepts of geometry come from the material
  properties of physical objects.  Practical experience with measuring
  instruments led to the abstract notions of Euclid.}
\label{fig:pythag}
\end{figure}

\section{Search for a Fundamental Geometry}

It is an ongoing study to find a fundamental geometry that might
include all of physics.  A very early goal of the author was to
discover an elementary but fully geometrical quantum mechanics.  This
was to be a simple incorporation of essential characteristics.  No
specific interpretation was presumed among the conflicting beliefs,
neither Copenhagen nor hidden variable, nor any other.  An elementary
understanding by geometry was expected to lead to reconciliation with
relativity and gravity.  Given the divergent metaphysical issues
within the subfields of physics, correctness was to be resolved by
recourse to experimental results without the interference of any
predisposed phenomenology.

It was soon realized that electrodynamics had to be included.  While
there are textbook examples of isolated quantum wave functions,
practical quantum effects are almost always observed by
electromagnetic interaction.  (The geometry has been found to support
this synthesis.)  However, a study involving radiation requires
further sophistication.  Because radiation is relativistic, special
relativity must be incorporated, a fact that is not so obvious in most
laboratory experiments.  The issues encountered in
quantum-electrodynamics become relevant.  Because of the usual
relativistic constraints, interaction potentials that depend on
space-like separated points must be relinquished in favor of covariant
forces acting along null lines.  Simple instructional examples are
hard to find.  Both the experimental phenomenology and the concomitant
geometrical representation are complicated.

\begin{figure}[ht]
\centerline{\psfig{figure=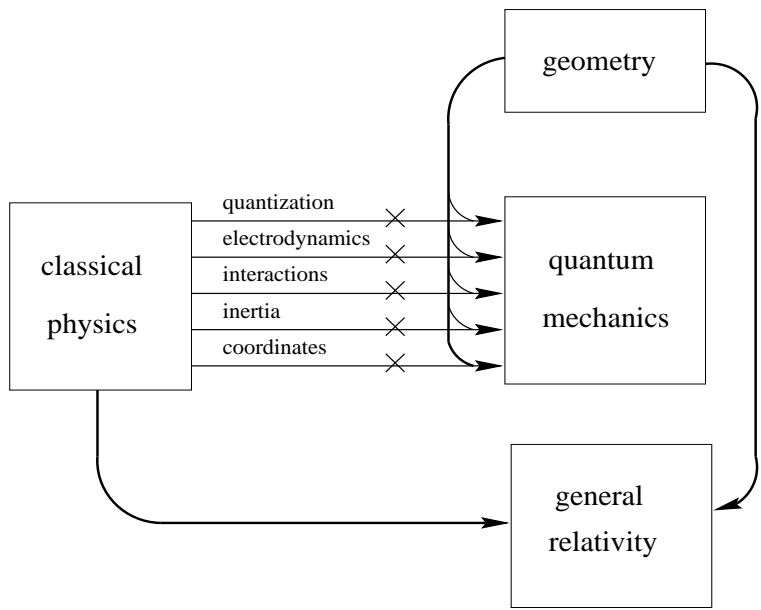,height=8cm,width=10cm}}
\caption{A geometrical theory displaces classical physics as a
fundamental theory and, in particular, as a progenitor of quantum
theory}
\label{fig:episto}
\end{figure}

It then became clear that a curvilinear description must be included
as well.  In general relativity the clock paradox is often used to
motivate the analogous development.  It is not so obvious that this
general type of difficulty occurs with other interactions.  A Lorentz
transformation is insufficient to describe any but the simplest motion
yet, geometrical equivalence must still be maintained.  While an
explicit physical ``clock'' paradox may not be available, neither
acceleration nor the complex motion of interference can be
modeled by linear coordinate transformations.  Any construction
suitable for quantum interference will require a curvilinear
system,  and a Riemannian formalism is necessary.  Realizing this
is essential to allow for the description of physical interactions in
a quantum-relativistic setting.

 These studies eventually developed into the aforementioned
quantum-geometrical theory.  To entertain  the question  of spinors, 
a selection of pertinent concepts is reviewed and the current results 
and conjectures are explained in this context.

\section{Epistemology and Geometrization}

  Attempts to relate quantum systems to gravitation
have shown unresolved difficulties.  The root causes originate in the
metaphysical assumptions of both subfields. New ideas are necessary
and some old ones may be untenable. 

 By convention, quantum mechanics is derived from classical mechanics
by quantization.  A classical Hamiltonian becomes a quantum wave
equation by the substitution of derivatives.  The process brings the
electrodynamic interaction across as well.  The geometrical sense of
coordinates is unchanged.  The partial differential equations of
quantum theory come into existence, but unfortunately without proper
evolution of concept.

In contrast to this, general relativity has some fundamental
geometrical basis outside of classical mechanics.  Herein is the
conflict.  Some physics has a geometrical component while some has
not.  Resolution of these differences is essential before
incorporating the intrinsically geometrical spinors.

As a progenitor of quantum mechanics, geometry may be capable of
displacing classical physics.  Early studies showed that some of the
essential quantum theory is buried in curvilinear geometrical
constructions.  It was apparent that formal quantization is not always
needed and that certain essential parts of quantum theory are found in
geometries unconverted by any quantization procedure.

A major change in the epistemological structure of physics is
proposed.  Classical physics is to be a simple phenomenology and
should not be used to generate quantum theory. The quantum is to come
from geometry alone.
  
This is realistic a possibility because there are no experiments that
favor classical over quantum. There are no observed classical point
particles and, without such particles, there can be no fundamental
basis for classical theory.  The active quantization of classical
mechanics allows inconsistencies to be incorporated.  Such
intellectual corruption is the root difficulty of a quantum theory of
gravity.  It is overly optimistic to believe that it is possible to
start with a fundamentally incorrect classical theory and arrive, by
any rigorous mathematical process, at a correct quantum theory.  The
referenced studies show that useful unified field theories can be
found once these changes are made to the origination of quantum laws.

\section{Quantization}

A review of the procedure for the conversion~\index{quantization} of a
classical theory to a quantum theory makes the limitations more
apparent.

The common  strategy is to start with an algebraic relation for 
the Hamiltonian such as,
\begin{equation}
H=\frac {p^2} { 2m} +V(x)
\label{eq:1}
\end{equation}
and convert it to
\begin{equation}
 -i \hbar \frac {\partial} {\partial t} = 
 \frac {1} {2m} \left( i \hbar \frac {\partial} { \partial x} \right)^2 +V(x),
\label{eq:2}
\end{equation}
by the substitution rule

\begin{align}
H & \rightarrow -i \hbar \frac {\partial} { \partial t}\\
p & \rightarrow i \hbar \frac {\partial} {\partial x}
 \label{eq:3}
\end{align}
The resulting equation, under restricted circumstances, is in
 agreement with experiment.  The mathematical process has been studied
 for some time, but never formally validated~\cite{quanto}.  

Attempting to quantize in a curvilinear system, makes plain the
enigmatic behavior.  Execution of the standard method 
requires a differential
substitution for energy and momentum.  The difficulty is that there
are already derivatives in the theory that lead to intrinsic 
non-commuting structures.
Specifically, the derivatives~\index{inconsistent derivatives} of the
metric tensor that are implicit in the covariance.  They persist in
the connections
\begin{equation}
 \Gamma^\alpha_{\beta \gamma} = { \frac {1} {2}} g^{\alpha \delta} 
\left( { \frac {\partial g_{ \delta \beta }} { \partial x^\gamma }}+ 
{\frac { \partial g_{\delta \gamma } } { \partial x^\beta }}  - 
{\frac { \partial g_{\beta \gamma}} { \partial x^\delta }} \right) .
\label{eq:4}
\end{equation}
which are integral to any curvilinear structure.  Such terms are
essential when there is a gravitational field.  A substitutionally
introduced derivative (an essential component of methodological quantization) 
may operate on a scalar, but because the operand
is not identified at the time of the substitution, combinations involving
tensors or spinors are problematical.
No resolution has become available and problems with quantum normal ordering
persist.  An accepted classical theory must contain the momentum $p$
and energy $E$ in such a way as to transform covariantly.  Because the
Christoffel symbols are required, the theory cannot be
quantized. Apparently, any ``correct'' classical gravitational theory
cannot be converted into a correct quantum theory by any formal
quantization.  The introduction of additional derivatives destroys the
covariance if even it can be defined unambiguously.

Moreover, the characteristic gravitational field equations are already second
order.  A quantization will increase the order, by at least one and
probably two.  The resulting equations are third or fourth order, even
neglecting complications generated by the non-locality of possible
expansions
of derivatives under the square root.  This is an epistemological
disaster because this accepted method to establish
a quantum-gravitational theory can no longer generate an intellectually
sound theory.  The alternate solution is preferred
here. Use covariant quantities to represent the wave function and then
generate field equations by careful selection of a geometrical system.

\section{Wave-particle Duality}

A deterministic and geometrical quantum theory impels the resolution
of wave-particle duality~\index{wave-particle duality}.  Conventional
wisdom says that real particles behave sometimes like particles and
sometimes like waves.  Double-talk not withstanding, the description
must be made more precise and the language adjusted to the reality of
experiments.

A precise notion of a classical  point particle is required and is
here taken to be an object that
is described by a one dimensional trajectory
(geodesic~\index{geodesic!classical}) in space-time.  There may be
additional parameters, such as mass or charge, that are part of the
dynamics. Such is the common notion.

A more sophisticated concept is required for the quantum particle.  A
one-dimensional trajectory is never sufficient.  Contributing to the
confusion is the accepted terminology that distinguishes a quantum
point objects as being non-composite.  Electrons are quantum point
like while protons or pions are not, this, even though none can be
condensed to a point.  The essential idea of a quantum point particle
is, in fact, simple cardinality. Electrons, by experiment, come in
sets that have a total number associated.  Point localization need not
occur.  The enumeration may be defined by weight, inertia, integrated
charge, or other interaction.  This notion of number~\index{particle
cardinality} is adequate for quantum mechanics.  It allows
normalization of the wave function.  Experimentally available
particles are countable but never
point-localized~\index{localization}.

The acceptance of quantization can subtly integrate the classical
idea of a point particle into the quantum interpretation.  The paradox
of wave particle duality then comes into play.  To illustrate,
consider the following diffraction experiment.

\begin{figure}[ht]
\centerline{\psfig{figure=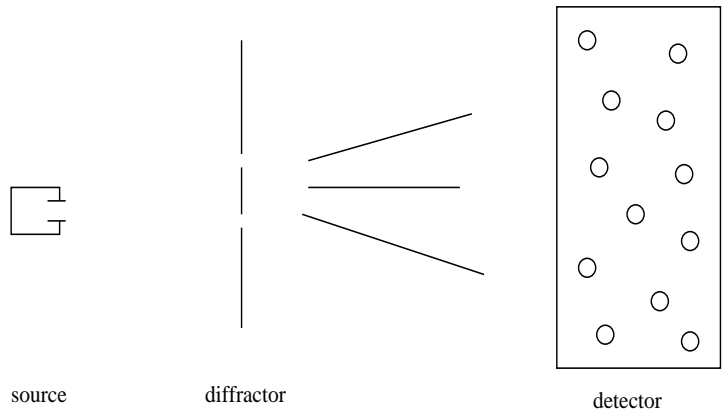,height=6cm,width=10cm}}
\caption{Diffracted electrons are captured in a detector.  An  explicit
model for the detector demonstrates the evolution of the 
wave function from an extended diverging wave to a compact bound state.}
\label{fig:detecto}
\end{figure}

As in the diagram, particles travel through a
diffraction~\index{diffraction} screen to be counted by a detector.
Suppose that the detector consists of a fixed uniform distribution of
charged centers, (perhaps protons), to which the electrons are
attracted. An individual electron will interact with these charged
objects, emit radiation, lose energy and ultimately be captured. Under
ideal conditions, it will bind to one of the charged centers.  The
large initial diffraction pattern will condense into an atomic wave
function.  Because of the multiple outcomes, this process generates
statistics.  The particular final center into which the electron
cascades is not predictable, but has a probability following the
quantum interpretation.

Within the geometrical theory, these statistics are not generated from
a fundamental supposition but are the result of the system evolution
as follows: (1) The collapse~\index{wave function collapse} of the
diffracted electron wave function proceeds in a finite time according
to the limits of relativity theory.  It is not instantaneous.  (2) The
forces which induce the process come from the advanced fields of the
particles that absorb the emitted photons.  The geometrical theories
require an explicit well-defined construction for these forces of
radiative reaction, otherwise equivalence cannot be guaranteed.  The
radiative reaction comes from ontological (electromagnetic) fields that
are derived from real particles.  (3) The radiation is unique to the
final electron state.  The equations are time reversible and
deterministic~\index{determinism}.  The possible final states of the
electron are complete and orthogonal.  Unitarity demands that the
final photon states map to the final electron state. (4) There is no
fundamental statistical mechanism.  The different results must be
taken as due to differences in the configuration of the system of
particles that absorb the photons as they are emitted.  Experiments
with radiation in cavities or those that study correlations make any
other conclusion hard to justify.

\begin{figure}[ht]
\centerline{\psfig{figure=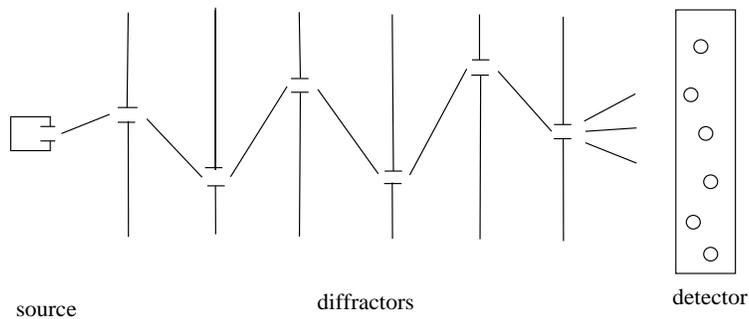,height=5cm,width=10cm}}
\caption{Successive diffractions appear to refine the particle position.  
In practice, this is always accompanied by interactions and radiation.}
\label{fig:diffeos}
\end{figure}

Suppose that the above experiment is extended by placing an additional
series of slits between the source and detector.  It is often argued
that such successive refinement, can identify the point position of a
particle.  Each slit diffracts the particle further, and by
integrating backwards along the probability current, a particular
trajectory within the initial emitted wave function can be defined to
any limit of precision.  It is sometimes argued that this shows that a
particle travels on a specific trajectory.  No interpretation of this
type is demanded by geometrical theory.  In fact, the backwards
projected refinement of the original trajectory is invalid because
radiative interactions are part of the selection process at each slit.

In consequence it is not experimentally possible to identify a
geometrical trajectory.  To describe particles as individual
but enumerable waves is sufficient, and this is entirely possible with
curvilinear geometry.  The wave function, as a collection of lines of
probability flow, rather than any single trajectory, is
associated with the geometry.

\section{Quantum Conformal Coupling}

The observation of a probability~\index{quantum-conformal coupling}
 involves counting particles as they are detected on a test screen.
 The experimental arrangement is controlled by material objects which
 serve as measuring rods to a local cartesian system.  In this sense 
the apparatus is fixed and
 unchanging.  A region on the screen is scribed to delineate an area
 in which particles will be counted as they are detected.
 The marked area is measured by dimensions
 $\delta x$ and $\delta y$ and the movement by
 $\delta z$ and $\delta t$ according to local metric $\dot g_{\mu
 \nu}$.  A count of particles, as given by quantum theory, is

$$N=\psi \psi^* \frac {\delta x \delta y \delta z}{ \delta t}$$

Consider now the observation of this effect by an observer using a
metric having a different conformal parameter~\index{conformal
parameter} and consequently a different scale, $\dot g^\prime_{\mu
\nu} = \lambda \dot g_{\mu \nu}$.  The new numerical marks are

\begin{equation}
(\delta x^\prime, \delta y^\prime, \delta z^\prime, \delta t^\prime)
= \sqrt{\lambda}\cdot  (\delta x, \delta y, \delta z, \delta t)
\label{eq:5}
\end{equation}

\begin{figure}[ht]
\centerline{\psfig{figure=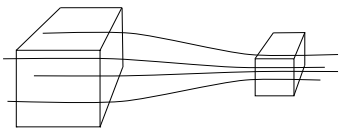,height=2cm,width=5cm}}
\caption{Local variations in the conformal parameter
of a particle space-time are equivalent to probability density waves.}
\label{fig:squeezo}
\end{figure}

The count becomes
\begin{equation}
N = \frac {\psi \psi^*} { \lambda} 
\frac {\delta x^\prime  \delta y^\prime \delta z^\prime}{ \delta t^\prime }
\label{eq:6}
\end{equation}
 and because $N$ is an invariant, $\psi$ must not be taken unchanged.
 A conformal transformation of the metric changes the measured
 probability density.  The wave function must be compensated.
 Presumably, $\sqrt{\lambda} \psi ^\prime = \psi$.  The conformal
 structure of the geometry cannot be separated from the quantum
 mechanical density.  It is proposed that as part of the 
physical-geometrical structure that many of the essential
 effects of quantum mechanics are manifestations of conformal
 structure~\index{conformal structure}.  A quantitative description
of particle localization is identified with the relative conformal
 density between the particle space and the observer.

The conformally invariant theories prevent this coupling and suppress
the intrinsic quantum characteristics of the geometry.  They cannot be
quantized because the geometrical density~\index{geometrical density}
is isolated from the quantum density~\index{quantum density}.  The
geometrical interpretation requires a coupling to the conformal
factor.  To avoid this problem,a single integrated quantum geometrical
conformal structure is used throughout this article.

\section{The Mechanical Gauge and the Physical Dimensionality}

Measurements of physical~\index{mechanical gauge} 
structure use clocks and rods whose
dimensional qualities originate outside of
general relativity.  Practically, the measuring standards come
from quantum mechanics.
\begin{figure}[ht]
\centerline{\psfig{figure=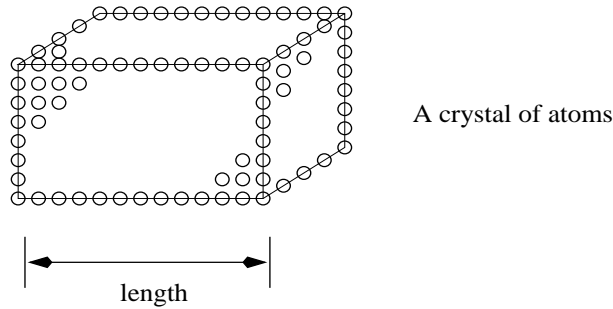,height=4cm,width=8cm}}
\caption{Macroscopic sizes and orientations are built up from the 
quantum properties of particles.}
\label{fig:crysto}
\end{figure}
However, without a consistent
quantum-gravitational theory, the inter-compatibility of this construction
is uncertain.   Quantum mechanics plays an essential role in determining the
geometrical size by referencing a fundamental length.  This
is taken as the Compton wavelength of the electron.
Enforcing a length standard removes the possibility of conformal
variations in the observer's metric.  It sets a particular gauge, here
called the mechanical gauge.  An experimentalist  assigns
the structure of space-time
in this way.  Objects rotate and do not change in length, at least 
in so far as it is possible to compare them  with each other.
Conformal stability of the observer's metric is enforced.
\begin{figure}[ht]
\centerline{\psfig{figure=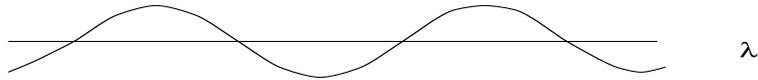,height=1cm,width=10cm}}
\caption{The fundamental local scale is referenced to the particles 
that make up the reference object.  The basic quantity 
is the Compton wavelength of the electron.}
\label{fig:waveo}
\end{figure}

In addition, the dimensionality of space-time is inferred by the
structure of mechanical objects.  These properties come from the
constituent particles.  In effect then, the three space coordinates
and one time coordinate, 3+1, are assigned by the properties of the
solutions of the quantum wave equation.  For the quantum gravitational
theory used here, the enveloping space is five-dimensional.  It is 
anticipated that a
higher number of dimensions may work and still provide an effective
space-time.  Such a larger geometrical construction may allow
interactions other than gravity or electromagnetism.

\section{Absolute Equivalence}

The equivalence of gravitational~\index{absolute equivalence} and
inertial mass has a long history.  Modern experiments are designed to
systematically test the relative contributions of different types of
mass-energy.  The integrated effects of electrodynamic, quantum, weak,
or strong forces are tested against each other with a sensitive balance.

\begin{figure}[ht]
\centerline{\psfig{figure=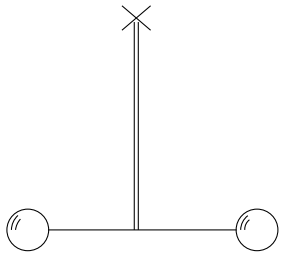,height=4cm,width=4cm}}
\caption{Equivalence is tested by balancing different materials in 
combined gravitational and accelerational fields.}
\label{fig:eveno}
\end{figure}

For gravity, exact equivalence is expressed by the
existence of a local frame.  The equivalence of other interactions 
apparently does not have such an elementary mathematical form of 
expression.  Yet, the experiments to date show that the integrated
equivalent of one type of mass is indistinguishable from  any
other.  In the general case, the coordinates that would be needed to
display the inter-transformation of forces are not experimentally
accessible.  Still the mathematical explanation of equivalence 
for all forces must have a formal structure that
goes beyond general relativity.  
Because current experiments show no discrepancies, some type of
equivalent theory is possible.

Here, exact absolute equivalence is assumed.  It is supposed that any
type of force can be transformed into any other type of force, at
least internally.  This assumption is easy to use, but is stronger
than experiments can establish directly.  The mechanical expression of
space-time, resulting in the reduced dimensionality, may conceal the
additional symmetries.  Verification must depend on the development 
of a specific geometrical thesis.

\begin{figure}[ht]
\centerline{\psfig{figure=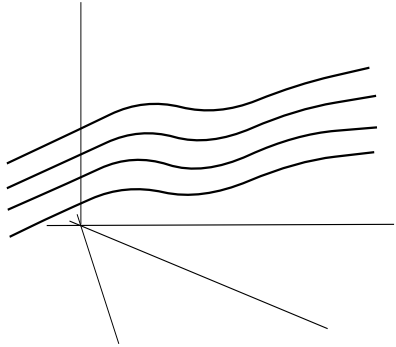,height=6cm,width=6cm}}
\caption{A quantum particle must be represented by all of the 
lines of its probability current.  A local inertial frame only 
applies to the local motion of a single quantum particle.  Each 
particle must have its own set of local frames.}
\label{fig:flowy}
\end{figure}

The principle of equivalence expressed as the existence of a local
frame can be adapted to a quantum particle.  Rather than using
an individual classical trajectory, the entire probability current 
of a wave-particle solution is taken at one time.  Only the 
local part of the quantum motion can be associated with a Lorentz frame.

Quantum-gravitational equivalence is more subtle than the equivalent
construction in classical
physics.  A convergent particle wave could be from either a
diffraction event or alternatively from the converging force of a
gravitational field. The equivalence of the wave equation
derivatives and the Christoffel derivatives is essential. 
Any identification of the difference requires knowledge of the cause
of the convergence.  It cannot be elucidated by study of the wave
alone.

\section{Essential Results in Five Dimensions}

For the conceptual development in hand, a mathematical summary of 
the five-dimensional~\index{dimensions!five}
 theory~\index{five dimensions}~\index{five-dimensional theory} is
 to the point.  Five coordinates 
$(x^0 \cdots x^4)\equiv x^m \sim (t,x,y,z,\tau)$ are
 chosen where $(x^0 \cdots x^3) \equiv x^\mu \sim (t,x,y,z,)$ can be taken to
 coincide with the observers space-time.  The five metric in standard
 form is
\begin{equation}
\gamma_{mn}=\left(
\begin{array}{cc}
  g_{\mu \nu} -  \A_\mu \A_\nu & \A_\mu \\
  \A_\nu & -1 \\
\end{array}
\right)
\label{eq:7}
\end{equation}
and can be rewritten with adjustable conformal factors as
\begin{equation}
\gamma_{mn}= \omega \left(
\begin{array} {cc}
\lambda g_{\mu \nu} - \chi^2 \A_\mu \A_\nu & \chi \A_\mu \\
 \chi \A_\nu & -1 \\
\end{array}
\right) .
\label{eq:8}
\end{equation}

There is a preferred set of geodesics~\index{quantum!geodesic}
\begin{equation}
\frac {dx^\mu} {ds} = \omega g^{\mu \nu} \chi \A_\nu
\label{eq:9}
\end{equation}
which are  tangent to the probability 
current~\index{geodesic!quantum}  and which
remain invariant under changes in $\lambda, \chi,$ and $ \omega$. 
It can be shown  that  these factors relate to the
causative explanation of the geodesic curvature in concordance 
with absolute equivalence.

A discussion of the use of $\chi$ and $\lambda$ is beyond this talk,
but $\omega$ is relevant.  With $\lambda
= 1$ and $\chi = 1$, the curvature scalar~\index{curvature scalar},
\index{scalar!curvature} $\Theta \equiv
\Theta(\gamma_{\mu \nu}) \equiv \Theta(\gamma_{\mu \nu}(\omega))$
becomes, upon being set equal to zero in a field free region,
\begin{equation}
(\partial)^2_{x^m}  \Psi \equiv 
\left( \frac {\partial^2}{\partial t^2} 
-\frac {\partial^2}{\partial x^2} 
-\frac {\partial^2}{\partial y^2} 
-\frac {\partial^2}{\partial z^2} 
-\frac{\partial^2}{\partial \tau ^2} \right)\Psi  =0
\label{eq:9b}
\end{equation}
Where the linear wave function $\Psi$ is equal to $\omega^{3/4}$.
Adding the condition~\index{mass}
\begin{equation}
\frac {\partial \Psi}{\partial \tau} = im 
\label{eq:10}
\end{equation}
 produces a  Klein-Gordon~\index{Klein-Gordon equation}  equation for mass $m$.
The quantum theory appears by the choice of geometry.

In fact, it turns out that the interactions can also be represented for 
another functional representation of the parameter$\omega$.
Setting to zero the Ricci tensor~\index{Ricci curvature!five-dimensional} 
  of the conformally multiplied metric
\begin{equation}
 0 = \Theta^{ij}( \gamma_{mn}(\omega)),
\label{eq:9a}
\end{equation}
it can be shown that this conformal contribution to the  
Ricci curvature  
of the original metric can be calculated as
\begin{equation}
\Theta^{ij}( \gamma_{mn}(0)) = T^{ij}(\omega).
\label{eq:11}
\end{equation}
with $T^{ij}$ known as a function of $\omega$.
And furthermore, for a suitable choice~\index{kappa@$\kappa$} of $\omega$,
this becomes, as expressed in four dimensional notation,
\begin{equation}
R^{\alpha \beta} = 8 \pi \kappa 
[F^\alpha_{\phantom{\alpha}\mu}F^{\mu \beta} 
+ m|\psi|^2 \frac {e^2}{m^2} \A^\alpha \A^\beta +
m|\psi|^2 \frac { 1-(e^2/m^2) \A^2}{ 2-(e^2/m^2)\A^2} g^{\alpha \beta}].
\label{eq:12}
\end{equation}
and at the same time~\index{alpha@$\alpha$}
\begin{equation}
F^{\beta \mu}|_\mu = 4 \pi \alpha  |\psi|^2 \A^\beta.
\label{eq:13}
\end{equation}

For spinors, the point is that the standard quantum field equation and the
effects of external interactions are mediated by the overall conformal
factor $\omega$.  In addition, the condition of inferred
dimensionality is satisfied.  The resulting system predicts
quantum-dynamical interactions that are in appearance
four-dimensional.  The mechanical gauge is correct and depends on the
separation constant, $m$, of the geometrical field equation.  In fact,
a universal mechanical gauge is realized by choosing the same type of
particle at each point and relating the scale of that mass to the
fifth coordinate.  (Scale changes in $\tau$ cause proportional changes
in the local mass of all particles and are unobservable.)

This construction allows for a theory of mass that comes from the
geometrical structure.  It is presumed that more complicated systems of
separation constants could be found in higher dimensional spaces and
so might give a mass spectrum.

\section{The Need for Spinors}

The motivation for the study of spinors is simply this: There are, at
present, no known stable massive bosons.  All decay into other
particles which, in the end, have spin $1/2$.  The internal mechanism
of the five theory does not directly accommodate spin.  None of the fields nor
the coordinates have spinor transformation properties.  Still group
theory predicts a deeper irreducible spinor representation of the
local coordinate transformations.  Apparently, spin is built into even
the simplest notion of geometry.  A full understanding remains elusive.

\begin{figure}[ht]
\centerline{\psfig{figure=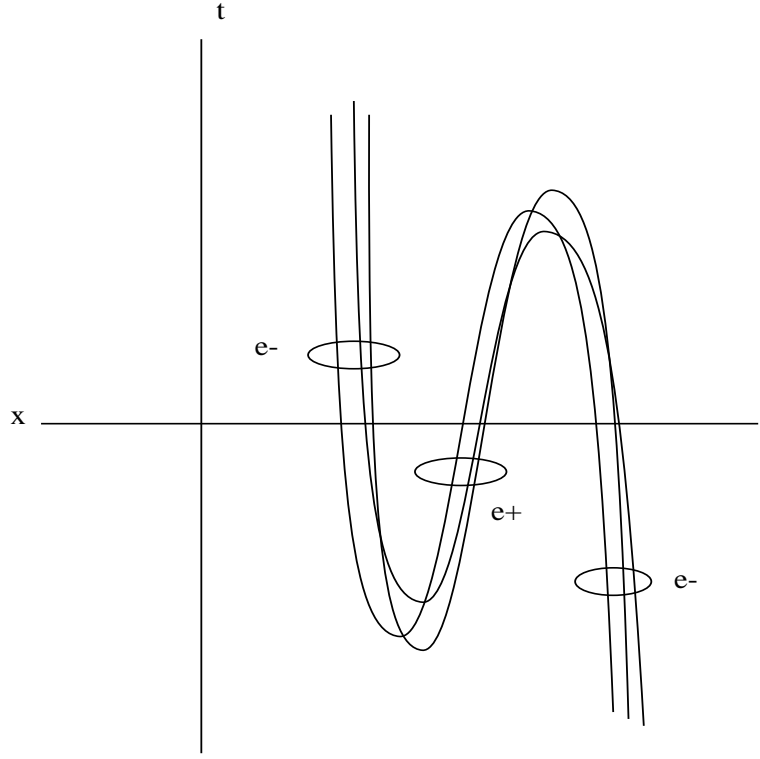,height=10cm,width=10cm}}
\caption{The positron from an e-p pair annihilates smoothly with 
another electron. The electrons must be  antisymmetric with 
respect to each other.}
\label{fig:twisty}
\end{figure}

It is known that the source current from a particular electron does
not interact with itself, but always with another.  Consider a pair
production event after which the positron anihiliates with an outside
electron.  While each electron must not be affected by its own source
currents, each must interact with the currents of the other even
though they eventually connect through the positron.  The enigma has
not been resolved in the context of a geometrical theory.
Antisymmetrization and the associated
indistinguishability~\index{particle indistinguishability} seem
important.  This and the ubiquitous presence of spin $1/2$ are taken
as symptoms of a deeper geometrical structure.

\section{Pauli-Dirac Theory}

The most elementary approach  is to reconsider the discussion of Pauli and
others \cite{pentaspin} in which the five anti-commuting Dirac 
matrices are each
associated with one of the five coordinate directions.
Let
\begin{equation}
\dot \gamma^0, \dot \gamma^1, \cdots, \dot \gamma^4 \equiv \dot
\gamma^m_{AB}
\label{eq:14}
\end{equation}
for $m=0, \cdots , 4$ and $ A,B = 1, \cdots , 4 $ such that
\begin{equation}
{ \bf 1} \dot \gamma^{mn} \equiv { \bf 1}
\left(
\begin{array}{cc}
 g^{\mu \nu} & 0 \\ 
 0 & -1 \\
\end{array}
\right) =
\frac {1}{ 2} \{ \dot \gamma^m , \dot \gamma^n \} \equiv
\frac {1}{ 2} (\gamma^m \gamma^n + \gamma^n \gamma^m)
\label{eq:15}
\end{equation}

It is convenient to use the  standard basis for $ \gamma^\mu$, $\mu=0
\cdots 3$ and to let $ \dot \gamma^4 = i \gamma^5 $.  In addition,
 define $\sigma^{mn} = [\gamma^m, \gamma^n] 
\equiv \gamma^m \gamma^n-\gamma^n \gamma^m$ and let  $\bf 1$ be the unit 
matrix.  For the
particle metric (as used to represent the fields of an individual
particle, say the electron.)
\begin{equation}
{\bf 1} \gamma^{mn} = {\bf 1}
\left(
\begin{array}{cc}
\dot g^{\mu \nu} & A^\mu \\ 
A^\nu & A^\tau A^\rho \dot g_{\tau \rho} -1 \\
\end{array}
\right)
 = \frac {1}{2} \{ \gamma^m , \gamma^n \}
\label{eq:16}
\end{equation}
where direct calculation gives $\gamma^\mu = \dot \gamma^\mu$ and
$\gamma^4 = \dot \gamma^4 - A_\mu \dot \gamma^\mu$

The standard Dirac equation
\begin{equation}
i \hbar \dot \gamma^\mu \frac {\partial}{\partial x^\mu} \psi -mc\psi =0
\label{eq:17}
\end{equation}
can be rewritten with the five dimensional notation.  Define a
similarity transformation by $S=\frac {1} {\sqrt{2}}({\bf 1}+\gamma^4)$, 
then left multiply by $S \dot \gamma^4$, right multiply by 
$e^{i\frac {mc}{ \hbar}x^4}$, and insert the pair $S^{-1} S$.
\begin{equation}
i\hbar ( S \gamma^4 \gamma^\mu S^{-1} 
\frac {\partial}{\partial x^\mu} S \psi 
e^{i \frac{mc}{\hbar} x^4} -mc S \gamma^4 S^{-1} \psi
e^{i \frac {mc}{\hbar} x^4} =0
\label{eq:18}
\end{equation}
Setting $\Psi = S \psi e^{i\frac {mc}{\hbar} x^4}$ and noting that
$S\gamma^4 S^{-1} = \gamma^4$ and $S\gamma^4 \gamma^\mu S^{-1} =
\gamma ^\mu$ gives
\begin{equation}
 \gamma^m \frac{\partial}{\partial x^m} \Psi =0.
\label{eq:19}
\end{equation}

An interesting question is how to connect the conformal part of $\Psi$
 with the five conformal factor $\omega$.  That is, if $\gamma^{ij}
 \rightarrow \omega \gamma^{ij}$, is this equivalent to $\gamma^i
 \rightarrow \omega^{1/2} \gamma^i$ or $ \Psi \rightarrow \omega^{1/2}
 \Psi$?  Certainly, there are conformal contributions to the curvature
 and other complexities as well, but because of the conformal flatness
 of the five theory, a construction of some kind should be possible.
 The conformal substructure of the Dirac theory is at issue.  A
 possible immediate application is a fundamental understanding of the
 interaction of spin one-half particles with the gravitational field.

The most interesting question is whether and in what way weak
interactions can be included.  Because experimental tests of
equivalence include the integrated rest-mass-energy of the weak
forces, a theory that can even describe weak interactions must show a
detailed equivalence if it is to be exact.  The internal gauge
transformation must exist. Perturbative descriptions of
neutrino-electron interactions should be the limit of exact
geometrical effects.  It is not yet entirely clear how to do this.
The five dimensional spinor approach is a direct avenue of
investigation.

\section{Spinor Coordinate Systems}

More radical is to search for a geometrical~\index{spinor coordinates}
system that has spinors as a natural tangent space.  For a classical
point particle, the choice of local frame is not unique.  Space
rotations about the particle center are degenerate.  For real
particles, the spin breaks the symmetry.  It is conventional to assign
a spinor basis space in place of a local Lorenz frame.  But, to
develop a fundamental geometry of spin, some sort of coordinates 
should themselves provide the local spinor orientation.  

\begin{figure}[ht]
\centerline{\psfig{figure=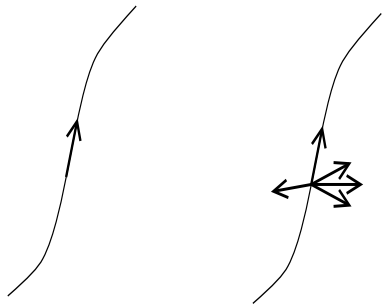,height=4cm,width=5cm}}
\caption{Electrons have spinor type local tangent frames.  These are not 
naturally derivable from the usual coordinate space.  A 
coordinate sub-manifold is sought.}
\label{fig:framo}
\end{figure}

Suppose that there is such a  base space.  Choose 
four complex valued coordinates~\index{complex coordinates} $
(\xi^1 \cdots \xi^4) \equiv \xi^A \in \BC$ which are to be related to
the physical coordinates $(x^0 \cdots x^4) \equiv x^m$.  Some of the
 usual difficulties with holonomy may not apply because of the conformal
flatness.

\begin{figure}[ht]
\centerline{\psfig{figure=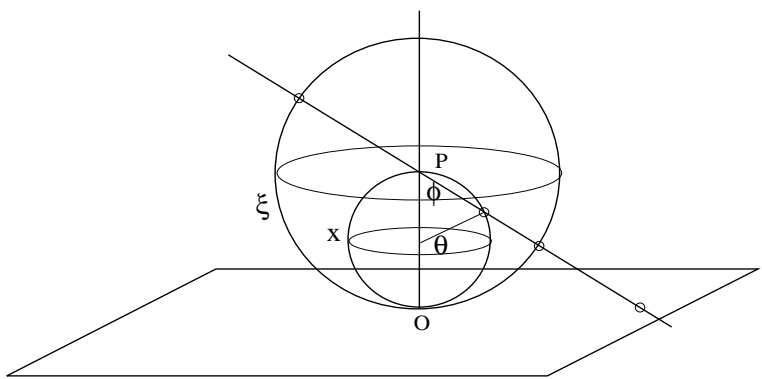,height=5.5cm,width=12cm}}
\caption{The complex plane can be mapped stereo-graphically to the surface 
of a sphere in two different ways.  The projection line from a 
point on the plane can go through either the center or the antipode.  
Both cases are shown together.  Two spheres of radius one and two 
 are tangent at the origin, point $ O$.  The
line  projection goes through the point $P$ which is at the center
of the larger sphere and at the antipode of the smaller one.
The angle $\theta$ from the vertical
axis to the projected point on the smaller sphere is exactly twice the
angle $\phi$ for the larger sphere.  There is a quadratic
mapping from one to the other.   The double valued relation of 
$\xi$ to $x$ is analogous to the required spinor map of $\BR^5$ to $\BC^4$
suggesting a spinor coordinate transformation.}
\label{fig:rollo}
\end{figure}

A number of transformations are possible, but not all have suitable
properties.   The following construction is under study.
Let  $\xi^A = \xi^A_r + i\xi^A_i $ with $\xi^A_r , \xi^A_i$ real.
 Define a coordinate relation by
\begin{equation}
 x^m = \frac {1}{2} \gamma^m _{AB} \overline \xi^A \xi ^B.
\label{eq:21}
\end{equation}
where $ \overline \xi^A $ is the complex conjugate
A five
dimensional real space is generated  by a restriction from the
four dimensional complex space.  
The important question is the  existence of differential 
equations and their relationship to more general coordinate 
transformations.

Further characteristics are useful. (1) Translations at least are
allowed in $\BC^4$ and also in $\BR^5$. (2) Apparently, expansion of the
equation $(\partial^2)_x \Psi = 0$ from five dimensions to the complex
four dimensional space gives something which may be written in the
form $(\partial^2)_\xi (\partial^2)_\xi \Psi = 0$ where
$(\partial^2)_\xi$ represents an eight dimensional d'Alembertian in
the $\xi_i^A$ and $\xi_r^A$.  The essential point is that the Klein
Gordon equation in five space can be written in terms of a suitable
differential invariant in spinor space.  Apparently, there are
characteristic solutions on $\BC^4$ that map to standard quantum wave
functions on $\BR^5$.  (3) The conformal structure of interactions in
5-space is transferred to the complex 4-space and induces a quantum
structure with the required gravito-electromagnetic interactions in
place.  (4) Also, it seems that the relevant interactions, as induced
by the second order derivatives of the conformal factor in $\BR^5$ are
augmented by other derivatives in $\BC^4$.  Thus additional conformal
variations, having nonzero derivatives for powers of $ \xi^A$ may not
all reduce to an expansion in derivatives by $x^m$.  These can be
neither electromagnetic, gravitational, nor quantum.  They are
conjectured to be the weak~\index{weak interactions} interactions.
The inferred electron-neutrino scattering should be analogous to the
conformal description of electron-photon and electron-graviton
effects.  (5) The five metric as an indefinite quadratic form is
generated from the the complex coordinate space,$\BC^4$, by the
properties of the gamma matrices.  The mechanical gauge comes from the
dynamical properties of particular solutions of equations on $\BC^4$
based on the symmetry of various combinations of coordinates as
specified by the $\gamma$'s.

A number of conjectures remain indeterminate.  (1) Is it possible to
represent the Dirac wave function as a unit vector in $\BC^4$
multiplied by a conformal factor?  (2) What, in detail, is the
differential geometry of this scheme?  (3) Is the phenomenology
correct?  (4) Can all know properties of the electron be described in
this way?  (5) What addition structures may be needed?  (6) Is it
correct to use general complex nonsingular transformations for the
$\xi$'s ?  (7) Are there other analogous coordinate
spaces and transformation systems that might apply to higher
particles?

\begin{figure}[ht]
\centerline{\psfig{figure=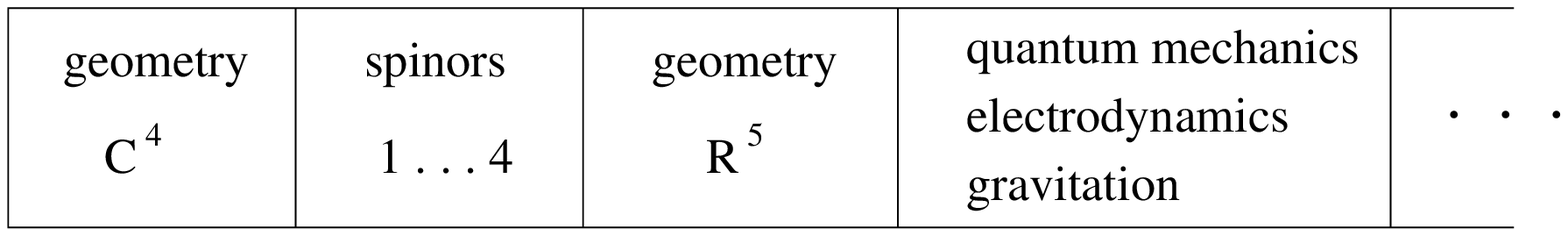,height=2cm,width=10cm}}
\caption{Geometry provides a basis for physics. 
The spinors may provide a physical-geometrical map 
between different dimensionalities.}
\label{fig:geomo}
\end{figure}

\section{Summary}

More than the presentation of any new result, this discussion is
intended to define an ongoing approach to an unsolved problem.  The
ultimate question, a geometry for all of physics, may find a solution
as new types of geometrical systems become associated with the known
experimental facts.  Certain ideas from the five dimensional
description of quantum mechanics and gravity enable the introduction
of spinors.  Spin has always been geometrical and its association with
other fundamental geometrical constructions should be fruitful.
Certain ideas are important.  Relinquishment of quantization abolishes
the need for a classical theory.  Exclusion of constructions that
depend on discreet classical particles allows a fully wave oriented
development.  A dynamics based on conformal effects permits a
mathematically simple scheme that is compatible with
spinors. Equivalence should be implied for all interactions.  The concept
of inferred dimensionality permits the use of spaces having more that
four dimensions to describe the physical-mechanical world.

Specific suggestions may bring the Dirac system into a geometrical
foundation.  The classical five-dimensional theory that associates
each one of the five anti-commuting Dirac matrices with one of the
five coordinates may be adaptable.  Explicit introduction of spinor
coordinates, particle by particle, may be possible.  In any case, a
fully geometrical structure with interactions is expected

The immediate goal, as a problem in physics, is to understand
electrons and their interactions.  The mathematical goal is to
understand how to handle these types of geometrical-particle
structures so that more complicated systems might be accessible in a
fully equivalent theory.

\normalsize
\subsection{Information about the Author}

\small
\vskip 1pc

{\obeylines
\noindent Daniel Galehouse
\noindent Department of Physics 
\noindent University of Akron
\noindent Akron, Ohio 44325
\noindent E-mail:dcg@uakron.edu
}
%\vskip 6pt
%\noindent Submitted: August 27, 2002 .\\

%\normalsize
%\noindent
%\printindex
\end{document}